\documentclass[twocolumn,prb,floatfix]{revtex4}
\usepackage{graphicx}    
\usepackage{dcolumn}     
\usepackage{bm}          
\usepackage{amsmath}          
\begin{document}
\preprint{Temperature}%

\title{Temperature dependence of spin-polarized transport
in ferromagnet / unconventional superconductor junctions}

\author{
T.~Hirai$^{1,2}$, Y.~Tanaka$^{1,2}$, N.~Yoshida$^{3}$, Y.~Asano$^{4}$,  
J.~Inoue$^{1}$ and S.~Kashiwaya$^{2,5}$
}%
%
\affiliation{
$^1$ Department of Applied Physics,
Nagoya University, Nagoya, 464-8603, Japan \\
$^2$ CREST Japan Science and Technology Cooperation (JST) 464-8603,
Japan \\
$^3$ Department of Microelectronics and Nanoscience, 
School of Physics and Engineering Physics, 
Chalmers University of Technology and Goteborg University, 
S-412 96 Goteborg, Sweden \\
$^4$ Department of Applied Physics, Hokkaido University, 
Sapporo 060-8628, Japan \\
$^5$ National Institute of Advanced Industrial Science and Technology, 
Tsukuba, 305-8568, Japan
}
%
\date{\today}
\begin{abstract}
Tunneling conductance 
in ferromagnet / unconventional superconductor junctions is
studied theoretically as a function of temperatures and 
spin-polarization in ferromagnets.
In $d$-wave superconductor junctions, a
zero-energy Andreev bound state drastically affects the temperature dependence 
of the zero-bias conductance (ZBC).
In $p$-wave superconductor junctions, 
numerical results show various temperature dependence of the ZBC
depending on the direction of the magnetic moment in ferromagnets
and the pairing symmetry in superconductors such as 
$p_{x}$, $p_{y}$ and $p_{x}+ip_{y}$-wave symmetries. The last one is  
a candidate for the pairing
 symmetry of Sr$_2$RuO$_4$.
%
%
From these characteristic features in the conductance,
we may obtain the information about 
the degree of spin-polarization in ferromagnets and
the direction of the $d$-vector in spin-triplet superconductors.
\end{abstract}
\maketitle
\section{Introduction}
In recent years, transport properties in unconventional superconductor 
junctions have been
studied both theoretically and experimentally.
In these junctions, a zero energy state (ZES) 
\cite{Buchholtz,Hara,Hu} formed at the junction interface 
plays an important role in the tunneling spectroscopy. 
It is now well known the ZES is responsible for zero-bias conductance peak (ZBCP)
 in the high-T$_C$ superconductor 
junctions~\cite{Kashiwaya1,Alff,Covington,Wei,Wang,Iguchi} 
and related phenomena~\cite{Tanaka96,T1,T2,asano,Tanu1,Tanu2,Tanu3,t1,t2,t3,t4,r1}. 
The theoretical studies clearly relate the formation of the ZES to the 
ZBCP in tunneling spectroscopy~\cite{Tanaka1,Kashiwaya2,Kashiwaya3,Lofwander}. 
Since the formation of the ZES is a general phenomenon in unconventional 
superconductor junctions, the
ZBCP is also expected in spin-triplet superconductor junctions~
\cite{Yamashiro1,Yamashiro2,Yamashiro3,Honerkamp,asano2,Kuroki3,Kusakabe}. 
Actually, the ZBCP has been observed in junctions of Sr$_2$RuO$_4$~\cite{Laube,Mao}
and UBe$_{13}$~\cite{Walti}. 
The ZBCP has also been theoretically predicted 
for organic superconductors (TMTSF)$_{2}$X very 
recently~
\cite{Sengupta,Kuroki1,Kuroki2}.

From a view of future device application, 
transport properties in hybrid structures consist of ferromagnets
and superconductors have attached much attention. 
It was pointed out   
in ferromagnet / insulator / spin-singlet unconventional superconductor 
(F/I/S) junctions that 
the amplitude of the ZBCP decreases with 
increasing the magnitude of the 
exchange potential in ferromagnets.
This is because the exchange potential breaks the time-reversal
symmetry and suppresses the retro-reflectivity of the 
Andreev reflection
~\cite{Andreev,Zhu,Kashiwaya4,Zutic,Yoshida,St,Hirai,Hirai1,Dong,YC2}. 
Thus the ZBCP is sensitive to the degree of spin-polarization in ferromagnets.
Since the tunneling conductance is independent of the 
magnitude of the insulating barrier~\cite{YoshidaC1}, 
it is possible to estimate the spin-polarization in ferromagnets 
through the temperature dependence of the ZBCP.
An experimental test would be carried out 
in La$_{0.7}$Sr$_{0.3}$MnO$_3$/YBa$_2$Cu$_3$O$_{7-x}$ junctions 
in near future~\cite{Chen,Sawa1,Sawa2,Yeh}.

When spin-triplet superconductors are attached to 
ferromagnets~\cite{Yoshida,St},
the ZBCP depends not only on the spin-polarization
also on other parameters such as relative angles between the $d$-vector in 
triplet superconductors 
and the magnetic moment in ferromagnets~\cite{Hirai,Hirai1,Kuroki3,Kusakabe}. 
Thus it may be possible to know
details of the pair potential by comparing 
the characteristic feature of the ZBCP in theoretical calculations
 and those in experiments.
For this purpose, it is necessary to know 
effects of another ingredients such as temperatures
and the profile of the pair potential near the junction interface on the ZBCP.
It is known that the amplitude of 
pair potential is drastically 
suppressed at a surface or a interface of superconductors in the 
presence of the ZES~\cite{Hara,Nagato1,Matsumoto,Ohashi,Yamashiro3,Tanuma2001}. 
Although there are several studies on tunneling 
phenomena in ferromagnet / unconventional superconductor junctions 
so far \cite{Buzdin,Sun}, 
such issues have never been addressed yet. 

In this paper, we calculate the 
tunneling conductance in ferromagnet / unconventional 
superconductor junctions as a function of temperatures
and degrees of spin-polarization in ferromagnets,  
where the spatial dependence of the 
pair potential is determined self-consistently 
based on the quasiclassical Green's function 
theory. 
We choose  $d$-wave, 
and $p_{x} + ip_{y}$-wave symmetries for the pair potentials 
which are candidates for pairing symmetries of high-$T_{C}$ cuprates
and Sr$_2$RuO$_4$, respectively. 
For comparison, 
we also study the conductance in $p_{x}$- and $p_{y}$-wave superconductor junctions.
From the calculated results, we reach the following conclusions. 

\noindent
(1) In $d$-wave junctions with (110) orientation and
$p_{x}$-wave junctions, an incident quasiparticle from a ferromagnet 
always feels the ZES irrespective of the incident angles. 
The zero-bias tunneling conductance (ZBC) at the zero temperature 
is insensitive to the barrier potential at the interface. 
This result indicates a possibility to estimate the 
magnitude of the spin-polarization of ferromagnets at sufficiently 
low temperatures ($T$) in experiments.

\noindent
(2)In $d$-wave junctions with (110) orientation and
$p_{x}$-wave junctions, the 
ZBC monotonically decreases with increasing temperatures 
for small magnitudes of spin-polarization. 
On the other hand for large magnitudes of polarization, 
the ZBC becomes an increasing function of temperatures. 
While for $d$-wave junctions with (100) orientation and
$p_{y}$-wave junctions, where the ZES does not appear,  
the ZBC is an increasing function of $T$ 
independent of the spin-polarization. 
For $p_{x}+ip_{y}$-wave junctions, 
the ZBC first decreases with increasing $T$ then increases. 

\noindent
(3)For $p$-wave junctions, the ZBC has various temperature dependence 
depending on the direction of the magnetic moment in 
ferromagnets. 
This unique property is peculiar to spin-triplet superconductors. 

\noindent
(4)
Throughout this paper, we calculate the ZBC in two ways; 
i) the spatial dependence of the pair potential is 
assumed to be the step function (non-SCF calculation), 
ii) the spatial depletion of the pair potentials is determined 
self-consistently (SCF-calculation). 
By comparing the conductance in the two ways,
we found that the results in the non-SCF calculation 
are qualitatively the same with those in the SCF-calculation. 

The organization of this paper is as follows. 
In Sec.~2, we formulate the tunneling conductance with arbitrary angle
between the magnetization axis of the ferromagnet and
$c$-axis of the superconductor. 
We show the tunneling conductance depends on the direction of the magnetic moments
 only when superconductors have spin-triplet pairing.
In Sec.~3, we calculate the polarization and temperature dependence of
ZBC for both $d$-wave and $p$-wave junctions.
In Sec.~4, we summarize this paper.

\section{Formulation}
\begin{figure}[hob]
\begin{center}
\includegraphics[width=10.0cm,clip]{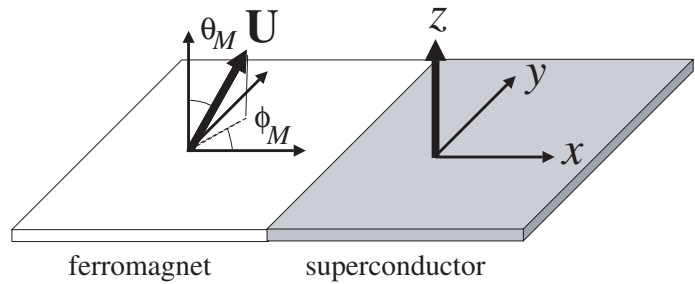}
\end{center}
\vskip -4mm
\caption{
Schematic illustration of a
ferromagnet / superconductor junction.
The direction of the magnetization axis is denoted by
a polar coordinate ($\theta_M$,$\phi_M$). 
\label{fig:1}
} 
\end{figure}

Let us consider a two-dimensional
F/I/S junction in the clean limit as shown in Fig.~\ref{fig:1}.
We assume a flat interface at $x=0$. The insulator is described 
by the delta-function $V(x)=H\delta (x)$, where $H$ 
represents the strength of the barrier potential.
We also assume that the Fermi energy $E_F$ and the effective mass $m$ in the ferromagnet
are equal to those in the superconductor.
The Stoner model is applied to describe ferromagnets, where
 the exchange potential $U$ characterizes the ferromagnetism.
The wave numbers in the ferromagnet for the
majority ($\bar{\uparrow}$) and the minority ($\bar{\downarrow}$) spins 
are denoted by
$k_{\bar{\uparrow} (\bar{\downarrow} )}
=\sqrt{\frac{2m}{\hbar^{2}}(E_{F} +(-) U)}$.
The wave functions $\Psi({\bf r})$ are obtained by solving the
Bogoliubov-de Gennes (BdG) equation under
the quasiclassical approximation~\cite{Millis,Bruder}
\begin{eqnarray}
E\Psi ({\bf r})=\int d{\bf r}'\tilde{H}({\bf r},{\bf r}')\Psi ({\bf r}')
\mbox{ , }
\Psi ({\bf r})=\left(
\begin{array}{c}
u_{\uparrow}({\bf r})\\ u_{\downarrow}({\bf r})\\
v_{\uparrow}({\bf r})\\ v_{\downarrow}({\bf r})
\end{array}
\right) \nonumber \\ 
\end{eqnarray}
\begin{eqnarray}
\tilde{H}({\bf r},{\bf r}')=\left(
\begin{array}{cc}
\hat{H}({\bf r})\delta ({\bf r}-{\bf r}') &
\hat{\Delta} ({\bf r} ,{\bf r'})\\
\hat{\Delta}^{\dag} ({\bf r},{\bf r}') &
-\hat{H}^{*}({\bf r})\delta ({\bf r}-{\bf r}')
\end{array}
\right),\nonumber \label{h1} 
\end{eqnarray}
where $E$ is the energy of a quasiparticle measured from $E_F$,
$\hat{H}({\bf r})
=h_0 \hat{\bf 1}-{\bf U}({\bf r})\cdot \boldsymbol{\sigma}({\bf r})$,
$h_0 =- \frac{\hbar^{2}}{2m}\nabla^{2} +V(x)-E_F$,
${\bf U}({\bf r})=U\Theta (-x){\bf n}$,
$\hat{\bf 1}$ and $\boldsymbol{\sigma}$ are
the ${2\times 2}$ identity matrix and the Pauli matrix, respectively. 
Here ${\bf n}$ points the direction of the magnetic moment in ferromagnets and
$\Theta (x)$ is the Heaviside step function.
The indices $\uparrow$ and $\downarrow$ denote the
spin degree of freedom of a quasiparticle in superconductors.
The magnetization axis in ferromagnets is represented 
in a polar coordinate $(\theta_M ,\phi_M)$ as shown in Fig. 1.
We assume that the quantization axis of spin in the 
triplet superconductors is in the $c$-axis which is parallel to $z$ direction.
At first, we assume that the pair potential is a constant
independent of $x$,
\begin{equation} 
\hat{\Delta}(\theta_S ,x)= \hat{\Delta}(\theta_S )\Theta (x),
\end{equation}
where $\hat{\Delta}(\theta_S )$ does not have the spatial dependence,
$k_x=k_F\cos\theta_S$ and $k_y=k_F\sin\theta_S$ are the wavenumber in
superconductors with $k_F$ being the Fermi wave number.
The pair potential $\hat{\Delta}(\theta_S )$
is given by
\begin{eqnarray}
\hat{\Delta}(\theta_S )=\left(
\begin{array}{cc}
\Delta_{\uparrow \uparrow}(\theta_S ) &
\Delta_{\uparrow \downarrow}(\theta_S )\\
\Delta_{\downarrow \uparrow}(\theta_S ) &
\Delta_{\downarrow \downarrow}(\theta_S )
\end{array}
\right).\label{dels-def}
\end{eqnarray}
The Hamiltonian in Eq.~(\ref{h1}) is written in the coordinate 
of the spin space in the superconductor. 
It is comprehensive to rewrite the Hamiltonian 
in the coordinate of spin space in the ferromagnet 
since this notation is useful to consider the scattering processes. 
The Hamiltonian in the coordinate of spin space in the ferromagnet 
is obtained by using the following unitary transformation:
\begin{align}
\tilde{H}_{F}({\bf r},{\bf r}')
=&\tilde{U}^{\dag}\tilde{H}({\bf r},{\bf r}')\tilde{U} \label{sut},\\
\tilde{U}=\left( 
\begin{array}{cc}
\hat{U} & 0 \\
0 & \hat{U}^{*}
\end{array}
\right),& \qquad
\hat{U}=\left( 
\begin{array}{cc}
\gamma_{1} & -\gamma_{2}^{*}\\
\gamma_{2} & \gamma_{1}^{*}
\end{array}
\right),\\
\gamma_{1} =\cos \frac{\theta_M}{2}e^{-i\phi_M},&
\qquad
\gamma_{2} =\sin \frac{\theta_M}{2}e^{i\phi_M},
\end{align}
where $\hat{U}$ is the operator which diagonalizes the
$\hat{H}({\bf r})$.
The effective pair potential in the coordinate of
spin space in ferromagnet is rewritten as
\begin{eqnarray}
\hat{\Delta}^F (\theta_S)
=\hat{U}^{\dag}\hat{\Delta} (\theta_S)\hat{U}^{*}.
\end{eqnarray}
Here, we consider the four types of pair potentials, 
$d$-wave, $p_{x}$-wave, $p_{y}$-wave 
and $p_{x}+ip_{y}$-wave symmetries in superconductors.
In $d$-wave case, the pair potential is described as
\begin{align}
\Delta_{\uparrow\downarrow}(\theta_S )
=& -\Delta_{\downarrow\uparrow}(\theta_S )\equiv 
\Delta_0 f(\theta_S), \\
\Delta_{\uparrow\uparrow}(\theta_S )=&\Delta_{\downarrow\downarrow}(\theta_S )
=0, \\
f(\theta_S) =& \cos [2(\theta_S -\alpha)], \label{f-defd}
\end{align}
where $\alpha$ is the angle between $a$-axis of the 
high-$T_c$ superconductors and the interface normal.
The effective pair potential in the coordinate of
spin space in ferromagnet is given by
\begin{align}
\hat{\Delta}^F (\theta_S )\equiv& 
\left( 
\begin{array}{cc}
\Delta_{\bar{\uparrow} \bar{\uparrow}}^{F}(\theta_S ) &
\Delta_{\bar{\uparrow} \bar{\downarrow}}^{F}(\theta_S )\\
\Delta_{\bar{\downarrow} \bar{\uparrow}}^{F}(\theta_S ) &
\Delta_{\bar{\downarrow} \bar{\downarrow}}^{F}(\theta_S )
\end{array}
\right), \\
=&\left(
\begin{array}{cc}
0 & \Delta_0 f(\theta_S)\\
-\Delta_0 f(\theta_S) & 0
\end{array}
\right), \\
=&\hat{\Delta}(\theta_S ).\label{tansn-d}
\end{align}
As shown in Eq.~(\ref{tansn-d}), the expression of the pair potential
remains unchanged under the transformation in Eq.~(\ref{sut}).
Therefore transport properties are expected to be independent of 
the direction of the magnetic moment. 
This conclusion can be applied to 
any spin-singlet superconductors.
On the other hand in spin-triplet superconductors, the pair potentials
are given by
\begin{align}
\Delta_{\uparrow\downarrow}(\theta_S ) =&\Delta_{\downarrow\uparrow}(\theta_S )
=\Delta_0 f({\theta_S}),\label{del-p1}\\
\Delta_{\uparrow\uparrow}(\theta_S ) =&\Delta_{\downarrow\downarrow}(\theta_S )
=0,\label{del-p2}
\end{align}
where the direction of the $d$-vector is parallel to the $c$-axis and
\begin{equation}
f({\theta_S}) = \left\{ \begin{array}{ccc} \cos\theta_S & \text{for} & p_{x}-\text{symmetry}, \\
                                           \sin\theta_S & \text{for} & p_{y}-\text{symmetry}, \\
 \textrm{e}^{i\theta_S} & \text{for} & p_{x}+ip_{y}-\text{symmetry}. 
\end{array}\right. \label{f-defp}
\end{equation}
Because the spin degree of freedom of Cooper pairs 
is active in triplet superconductors,  
the pair potential after the transformation in Eq.~(\ref{sut}) depends 
on the direction of the magnetic moment
\begin{eqnarray}
\hat{\Delta}^F (\theta_S )=\left( 
\begin{array}{cc}
\sin \theta_M & \cos \theta_M \\
\cos \theta_M & -\sin \theta_M
\end{array}
\right) f(\theta_S)\Delta_{0}. \label{del_f}
\end{eqnarray}
There are four reflection processes
when an electron with the majority spin is incident from ferromagnets:\\
i) Andreev reflection to majority spin
($a_{\bar{\uparrow}\bar{\uparrow}}$)\\
ii) Andreev reflection to minority spin
($a_{\bar{\uparrow}\bar{\downarrow}}$)\\
iii) normal reflection to majority spin
($b_{\bar{\uparrow}\bar{\uparrow}}$) and\\
iv) normal reflection to minority spin
($b_{\bar{\uparrow}\bar{\downarrow}}$).\\
Similar reflection processes are also possible,
 when an electron with the minority spin is incident from ferromagnets.
The Andreev and the normal reflection coefficients
are denoted by $a_{\bar{s}\bar{s}'}$ and $b_{\bar{s}\bar{s}'}$,
respectively.
In these coefficients, a quasiparticle is reflected from the spin-channel $\bar{s}$ 
into the spin-channel $\bar{s}'$.

The wave function in ferromagnets for majority spin injection is represented by
\begin{eqnarray}
\Psi_{\bar{\uparrow}}(x)
&=&e^{ik_{F\bar{\uparrow}}x}
\left( 
\begin{array}{c}
1\\0\\0\\0
\end{array}
\right)
+a_{\bar{\uparrow}\bar{\uparrow}}e^{ik_{F\bar{\uparrow}}x}
\left(
\begin{array}{c}
0\\0\\1\\0
\end{array}
\right)
+a_{\bar{\uparrow}\bar{\downarrow}}e^{ik_{F\bar{\downarrow}}x}
\nonumber \\ 
&\times&
\left(
\begin{array}{c}
0\\0\\0\\1
\end{array}
\right) 
+b_{\bar{\uparrow}\bar{\uparrow}}e^{-ik_{F\bar{\uparrow}}x}
\left(
\begin{array}{c}
1\\0\\0\\0
\end{array}
\right)
+b_{\bar{\uparrow}\bar{\downarrow}}e^{-ik_{F\bar{\downarrow}}x}
\left(
\begin{array}{c}
0\\1\\0\\0
\end{array}
\right),\nonumber \\ 
\end{eqnarray}
where $k_{F\bar{\downarrow}}<k_S <k_{F\bar{\uparrow}}$ and
$k_S \approx \sqrt{\frac{2mE_F}{\hbar^2}}$.
The wave function for minority spin injection
is written in the similar way.
The coefficients
$a_{\bar{s}\bar{s}'}$ and $b_{\bar{s}\bar{s}'}$ are
determined by solving the BdG equation
with the quasiclassical approximation
under appropriate boundary conditions.

The tunneling conductance $\sigma_{T}(eV)$ for finite temperature
is given by \cite{BTK,Zaitsev,Shelankov}
\begin{align}
\sigma_{T}(eV)&=\frac{2 e^{2}}{h} G,\\
G 
&=\frac{1}{16k_{B}T}
\int_{-\infty}^{\infty}dE
\int_{-\pi /2}^{\pi /2}d\theta_S \cos{\theta_S} \nonumber \\
 \times &
\left( \sigma_{S\bar{\uparrow}}(\theta_S)
+\sigma_{S\bar{\downarrow}}(\theta_S)
\right) 
\mbox{sech}^2 \left( \frac{E-eV}{2k_B T}\right),
\end{align}
\begin{align}
\sigma_{S\bar{\uparrow}}=&1+|a_{\bar{\uparrow}\bar{\uparrow}}|^2
-|b_{\bar{\uparrow}\bar{\uparrow}}|^2
+\left( \frac{\eta_{\bar{\downarrow}}}{\eta_{\bar{\uparrow}}}
|a_{\bar{\uparrow}\bar{\downarrow}}|^2
-\frac{\eta_{\bar{\downarrow}}}{\eta_{\bar{\uparrow}}}
|b_{\bar{\uparrow}\bar{\downarrow}}|^2\right) \nonumber\\
&\times \Theta (\theta_C -|\theta_S |),\\
\sigma_{S\bar{\downarrow}}=&\left(
1+\frac{\eta_{\bar{\uparrow}}}{\eta_{\bar{\downarrow}}}
|a_{\bar{\downarrow}\bar{\uparrow}}|^2
+|a_{\bar{\downarrow}\bar{\downarrow}}|^2
-\frac{\eta_{\bar{\uparrow}}}{\eta_{\bar{\downarrow}}}
|b_{\bar{\downarrow}\bar{\uparrow}}|^2
-|b_{\bar{\downarrow}\bar{\downarrow}}|^2 \right)\nonumber\\
&\times  \Theta (\theta_C -|\theta_S |),
\end{align}
with $Z_{\theta_S}=Z/\cos{\theta_S}$, $Z=2mH/\hbar^2 k_F$ and
$\eta_{\bar{\uparrow} (\bar{\downarrow} )}=
\sqrt{1\pm X /\cos^2 \theta_S}$.
Here $X=U/E_F$ is defined as the spin-polarization parameter.
The quantity $\sigma_{S\bar{\uparrow} (\bar{\downarrow} )}$
is the tunneling conductance for an incident electron 
with the majority (minority) spin.
For $|\theta_S |>\theta_C =\cos^{-1}\sqrt{X}$,
the reflected wave becomes an evanescent wave and does not 
contribute to the tunneling conductance.
As shown in above equations, 
the tunneling conductance depends on $\theta_M$ only
when superconductors have spin-triplet Cooper pairs.
The tunneling conductance can be summarized in 
simple equations in following several cases.
When $\theta_M$ is 0 or $\pi$ in spin-triplet superconductors,
the conductance is described by
\begin{align}
\sigma_{S\bar{\uparrow}}=&
\sigma_{N\bar{\uparrow}} (A +  B), \label{sig-1} \\
\sigma_{S\bar{\downarrow}}=&\sigma_{N\bar{\downarrow}} C
\end{align}
\begin{align}
A=& [1-|\Gamma_+ \Gamma_- |^2 (1-\sigma_{N\hat{\downarrow}})
+\sigma_{N\bar{\downarrow}}|\Gamma_+|^2] \nonumber \\
&\times \Theta (\theta_C -|\theta_S |)/L_{D1}, \\
B=& 1-\Theta (\theta_C -|\theta_S |)
[1 - \mid \Gamma_{+}\Gamma_{-} \mid^{2}]/L_{D1}, \\
C=&
[1-|\Gamma_+ \Gamma_- |^2 (1-\sigma_{N\bar{\uparrow}})
+\sigma_{N\bar{\uparrow}}|\Gamma_+|^2]\nonumber\\
&\times \Theta (\theta_C -|\theta_S |)/L_{D2},\\
L_{D1}=&\left| 1-\Gamma_+ \Gamma_- \sqrt{1-\sigma_{N\bar{\downarrow}}}
\sqrt{1-\sigma_{N\bar{\uparrow}}} \right.\nonumber \\
&\times\left. 
\exp\left[ i(\varphi_{\bar{\downarrow}}-\varphi_{\bar{\uparrow}})\right] \right|^2, \\
L_{D2}=&\left| 1-\Gamma_+ \Gamma_- \sqrt{1-\sigma_{N\bar{\downarrow}}}
\sqrt{1-\sigma_{N\bar{\uparrow}}} \right.\nonumber \\
&\times\left. 
\exp\left[ i(\varphi_{\bar{\uparrow}}-\varphi_{\bar{\downarrow}})\right] \right|^2,
\end{align}
with
\begin{align}
\exp(i\varphi_{\bar{\downarrow}})=&
\frac{1-\eta_{\bar{\downarrow}}+iZ_{\theta_S}}
{\sqrt{1-\sigma_{N\bar{\downarrow}}}
(1+\eta_{\bar{\downarrow}}-iZ_{\theta_S})},
\\
\exp(-i\varphi_{\bar{\uparrow}})=&
\frac{1-\eta_{\bar{\uparrow}}-iZ_{\theta_S}}
{\sqrt{1-\sigma_{N\bar{\uparrow}}}
(1+\eta_{\bar{\uparrow}}-iZ_{\theta_S})}.
\end{align}
Above equations can be applied to the spin-singlet superconductors.
When $\theta_M =\pi /2$ in spin-triplet superconductors,
the conductance for each spin is given by
\begin{align}
\sigma_{S\bar{\uparrow}}=&\sigma_{N\bar{\uparrow}}
\frac{1-|\Gamma_+ \Gamma_- |^2(1-\sigma_{N\bar{\uparrow}})
+\sigma_{N\bar{\uparrow}}|\Gamma_+ |^2}
{|1-\Gamma_+ \Gamma_- (1-\sigma_{N\bar{\uparrow}})|^2},
\\
\sigma_{S\bar{\downarrow}}=&\sigma_{N\bar{\downarrow}}
\frac{1-|\Gamma_+ \Gamma_- |^2(1-\sigma_{N\bar{\downarrow}})
+\sigma_{N\bar{\downarrow}}|\Gamma_+ |^2}
{|1-\Gamma_+ \Gamma_- (1-\sigma_{N\bar{\downarrow}})|^2} \nonumber \\
& \times
\Theta (\theta_C -|\theta_S |).
\end{align}
In above equations, we define 
\begin{align}
\Gamma_{\pm}=&\pm \frac{E-\sqrt{E^2-|\Delta (\theta_S )|^2}}
{\Delta (\theta_S)^{*}}, \label{def-gam}\\
\sigma_{N\bar{\uparrow}}=&\frac{4\eta_{\bar{\uparrow}}}
{(1+\eta_{\bar{\uparrow}})^2 +Z_{\theta_S}^{2}},\\
\sigma_{N\bar{\downarrow}}=&\frac{4\eta_{\bar{\downarrow}}}
{(1+\eta_{\bar{\downarrow}})^2 +Z_{\theta_S}^{2}}
\Theta (\theta_C -|\theta_S |).
\end{align}

In this paper, the dependence of the pair potential on temperatures
is described by the BCS's gap equation.
The spatial dependence of the pair potential 
can be described by 
$\hat{\Delta}(\theta_{S},x)$ 
with $\hat{\Delta}(\theta_{S},x)
=\check{\Delta}(\theta_{S},x)\Theta(x)$.
In order to determine the spatial dependence of 
$\check{\Delta}(\theta_{S},x)$, we apply the 
 quasiclassical Green's function theory
developed by Hara, Nagai, et. al.~
\cite{Hara,Ashida,Nagato,Ohashi}. 
In the following, we briefly explain the method in the case of 
spin-singlet superconductors. An extension to spin-triplet 
superconductors is straightforward. 
The spatial dependence of $\check{\Delta}(\theta_{S},x)$ 
is calculated by the diagonal elements of the matrix 
Green function $g_{\alpha\alpha}(x)$ which 
are represented by 
\begin{align}
g_{++}(\theta_S ,x)=&i
\left( 
\begin{array}{cc}
\frac{1+D_{+}(x)F_{+}(x)}{1-D_{+}(x)F_{+}(x)}   
& \frac{2iF_{+}(x)}{1-D_{+}(x)F_{+}(x)}\\
\frac{2iD_{+}(x)}{1-D_{+}(x)F_{+}(x)}
 & -\frac{1+D_{+}(x)F_{+}(x)}{1-D_{+}(x)F_{+}(x)}   
\end{array}
\right),\label{gpp}\\
g_{--}(\theta_S ,x)=&i
\left( 
\begin{array}{cc}
\frac{1+D_{-}(x)F_{-}(x)}{-1+D_{-}(x)F_{-}(x)}   
& \frac{2iF_{-}(x)}{-1+D_{-}(x)F_{-}(x)}\\
\frac{2iD_{-}(x)}{-1+D_{-}(x)F_{-}(x)}
 & -\frac{1+D_{-}(x)F_{-}(x)}{-1+D_{-}(x)F_{-}(x)}   
\end{array}
\right)\label{gmm},
\end{align}
where an index $\alpha =\pm$ 
specifies the direction of the momentum  
in the $x$ direction.
In these Green functions, $D_{\alpha}(x)$ and $F_{\alpha}(x)$ 
obey the following equations
\begin{align}
&\hbar |v_{Fx}|D_{\alpha}(x)= \alpha \nonumber\\
& \times \left[2\omega_m D_{\alpha}(x)+\bar{\Delta }(\theta_S ,x)D_{\alpha}^{2}(x)
-\bar{\Delta}^{*}(\theta_S, x)\right],\label{df1}\\
&\hbar |v_{Fx}|F_{\alpha}(x)
=\alpha \nonumber\\
&\times \left[ 
-2\omega_m F_{\alpha}(x)+\bar{\Delta}^{*} (\theta_S ,x)F_{\alpha}^{2}(x)
-\bar{\Delta}(\theta_S, x)\right],\label{df2}\\
&\check{\Delta}(\theta_S, x)=
\left( 
\begin{array}{cc}
0 & \bar{\Delta}(\theta_{S},x) \\
-\bar{\Delta}(\theta_{S},x)  & 0 
\end{array}
\right).
\end{align}
The boundary conditions at the interface are given by 
\cite{Tanuma2001,Hirai2}
\begin{align}
F_{+}(0)=&\frac{(\eta_{\bar{\uparrow}}-1+iZ)
(\eta_{\bar{\downarrow}}-1-iZ)}
{(\eta_{\bar{\uparrow}}+1+iZ)
(\eta_{\bar{\downarrow}}+1-iZ)}
D_{-}(0)^{-1},\\ 
F_{-}^{-1}(0)=&\frac{(\eta_{\bar{\uparrow}}-1+iZ)
(\eta_{\bar{\downarrow}}-1-iZ)}
{(\eta_{\bar{\uparrow}}+1+iZ)
(\eta_{\bar{\downarrow}}+1-iZ)}
D_{+}(0).
\end{align}
We first solve $D_{\pm}(x)$ and $F_{\pm}(x)$ in Eqs.~(\ref{df1}) and (\ref{df2}), 
then calculate $g_{\pm,\pm}(\theta,x)$ in Eqs.~(\ref{gpp}) and (\ref{gmm}) 
for a given $\bar{\Delta}(\theta_{S},x)$. 
By using $g_{\pm,\pm}(\theta,x)$,  
$\bar{\Delta}(\theta_{S},x)$ is given by the 
following equations 
\begin{align}
\check{\Delta}(\theta_{S},x)
=&  \sum_{n \alpha} \int^{\pi/2}_{\pi/2}
\;d\theta_{S'} V(\theta_{S},\theta_{S}')
g_{\alpha,\alpha}(\theta_{S}',x),\\
V(\theta,\theta') =&g_{0}f(\theta)f(\theta'),\\
g_{0}=&
\frac{2\pi k_{B}T}{{\rm ln}(T/T_{C})
+\sum_{0<m<m_{a}}
\frac{1}{m + 1/2}  },
\end{align}
where we introduce the cut-off $m_{a}$ to regularize $g_0$ and  
$f(\theta)$ is given in  Eqs.~(\ref{f-defd}) and (\ref{f-defp}).
The iteration is carried out until the sufficient convergence 
is obtained. In this way, we obtain $\bar{\Delta}(\theta,x)$, i.e., 
$\hat{\Delta}(\theta,x)$ self-consistently. 

Under the pair potential in the self-consistent calculation, 
we obtain $\check{\Gamma}_{\pm}(x)$ by solving 
\begin{align}
&i \hbar |v_{Fx}|\check{\Gamma}_{+}(x) \nonumber \\ 
=&\alpha \left[ 
2E \check{\Gamma}_{+}(x) 
- \bar{\Delta} (\theta_S ,x)\check{\Gamma}_{+}^{2}(x)
-\bar{\Delta}^{*}(\theta_{S}, x)\right],\\ 
& i \hbar |v_{Fx}|\check{\Gamma}_{-}(x) \nonumber \\
=&\alpha \left[ 
2E \check{\Gamma}_{-}(x) 
- \bar{\Delta}^{*}(\theta_{S}, x) \check{\Gamma}_{-}^{2}(x) 
-\bar{\Delta}(\theta_{S}, x)\right]. 
\end{align}
By substituting $\Gamma_{\pm}$ into Eqs.~(\ref{sig-1})-(\ref{def-gam}), 
we can calculate the tunneling conductance. 
In what follows, we assume that the transition temperature
of ferromagnets is much larger than $T_{C}$ which is the 
transition temperature of superconductors. 
In such situation, we can neglect the temperature dependence of $X$. 

\section{Results}

\subsection{polarization dependence of zero-bias conductance}
In this subsection, we show calculated results of 
the ZBC at the zero temperature as a function of the spin-polarization in 
ferromagnets ($X$).
The dimensionless ZBC ($\Gamma$) is given by  
\begin{equation}
\Gamma = \frac{ \sigma_{T}(0) h}{2e^{2}}.
\end{equation}
At first we show the conductance obtained in the step-function model, 
where the spatial dependence of the pair potential is not determined self-consistently 
(non-SCF calculation). The $X$-dependence of $\Gamma$ at the zero temperature for 
$d$-wave junctions is plotted in Fig.~2.
The magnitude of $\Gamma$ is always a decreasing function of $X$. 
For $\alpha =0$ (Fig.~2(a)), $\Gamma$ decreases with increasing of $Z$.
On the other hand, for $\alpha =\pi /4$ (Fig. 2(b)),
$\Gamma$ is completely independent of $Z$
because of the perfect Andreev reflection due to 
the zero-energy resonance state at the interface.
\begin{figure}[htb]
\begin{center}
\scalebox{0.4}{
\includegraphics[width=10.0cm,clip]{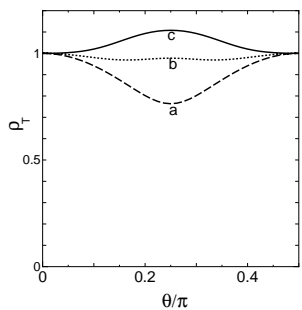}
}
\caption{
$X$ dependence of the zero-bias conductance
$\Gamma$ in non-SCF calculation
for (a) $\alpha =0$ and (b) $\alpha =\pi /4$
in $d$-wave junctions at zero temperature.
a: $Z=0$, b: $Z=1$ and c: $Z=5$.
 }
\end{center}
\end{figure}

Secondly, we show the polarization dependence of $\Gamma$ 
in triplet $p_{x}$, $p_{y}$ and $p_{x}+ip_{y}$-wave 
superconductor junctions as shown in Fig.~3, 
where the pair potential are given in 
Eqs.~(\ref{del-p1}), (\ref{del-p2}) and (\ref{f-defp}).
The conductance depends on $\theta_M$ for all pairing
symmetries. The spin degree of
freedom remains in spin-triplet superconductors.
As a consequence, the conductance depends on the relative angle between
the magnetic moment in ferromagnets and $d$-vector in superconductor.
This is the characteristic feature of $p$-wave junctions. 
For $\theta_M =0$, as shown in Figs.~3 (a), (d), and (g),
$\Gamma$ approaches to zero in the limit of
$X \rightarrow 1$ independent of $Z$. 
In these cases, the diagonal elements in Eq.~(\ref{dels-def}) disappear 
and a quasiparticle suffers the spin-flip in the Andreev reflection,
(i.e., $a_{\bar{\uparrow},\bar{\uparrow}} =
a_{\bar{\downarrow},\bar{\downarrow}} = 0$). 
For $|\theta_S | > \theta_C$,
the Andreev reflection to $\bar{\downarrow}$ spin 
becomes the evanescent wave. 
At the same time, an incident wave with $\bar{\downarrow}$ spin vanishes. 
Thus $\Gamma$ vanishes in the limit of $X=1$,
where ferromagnets are referred to as half-metals.
On the other hand, for $\sin\theta_M \neq 0$,
$\Gamma$ takes finite values even in $X \rightarrow 1$
as shown in 
Figs.~3 (b) (c), (e) (f)  (h), and (i)
 because the spin-conserved Andreev reflection is still
possible in these junctions.
The results for $\theta_M =\pi /2$ are shown in 
Figs.~3 (c) (f) and (i).
In this case, the off-diagonal elements in Eq.~(\ref{del_f}) become zero.
Thus spin of a quasiparticle is conserved in the Andreev reflection,
(i.e., $a_{\bar{\uparrow},\bar{\downarrow}} =
a_{\bar{\uparrow},\bar{\downarrow}} = 0$). 
The Andreev reflection of an electron with $\bar{\uparrow}$ spin survives
irrespective of $\theta_S$, whereas that of an electron with $\bar{\downarrow}$ 
spin vanishes for $|\theta_S |  > \theta_C$.
We note in $p_{x}$-wave junctions that $\Gamma$ does not depend on $Z$
as well as $d$-wave junctions with $\alpha =\pi /4$.
This is because the ZES's are formed at the interface.  

\begin{figure}[htb]
\begin{center}
\scalebox{0.4}{
\includegraphics[width=15.0cm,clip]{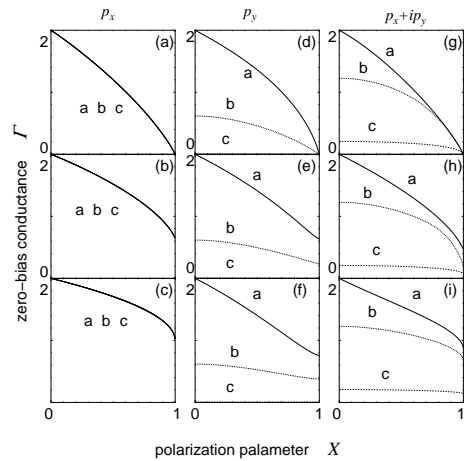}
}
\caption{
$X$ dependence of $\Gamma$ in non-SCF calculation.
(a) $\theta_M =0$, (b) $\theta_M =\pi /4$ 
and (c) $\theta_M =\pi /2$ for $p_x$-wave junctions. 
(d) $\theta_M =0$, (e) $\theta_M =\pi /4$
and (f) $\theta_M =\pi /2$ for $p_y$-wave junctions. 
(g) $\theta_M =0$, (h) $\theta_M =\pi /4$
and (i) $\theta_M =\pi /2$ for $p_x +ip_y$-wave junctions. 
a: $Z=0$, b: $Z=1$ and c: $Z=5$. } 
\end{center}
\end{figure}

Thirdly  we show the tunneling conductance
under the pair potential whose spatial dependence is determined 
self-consistently ( SCF calculation). 
The results for $d$, $p_{x}$, $p_{y}$, and $p_{x}+ip_{y}$-wave junctions 
are shown in 
Figs.~4 and 5.
The conductance in SCF calculation in 
Figs.4 and 5 
should be compared with corresponding results in non-SCF calculation in 
Figs.~2 and 3, respectively.
We do not find any remarkable differences between the results 
in SCF calculation
and those in non-SCF calculation as shown in these figures.
\begin{figure}[htb]
\begin{center}
\scalebox{0.4}{
\includegraphics[width=12.0cm,clip]{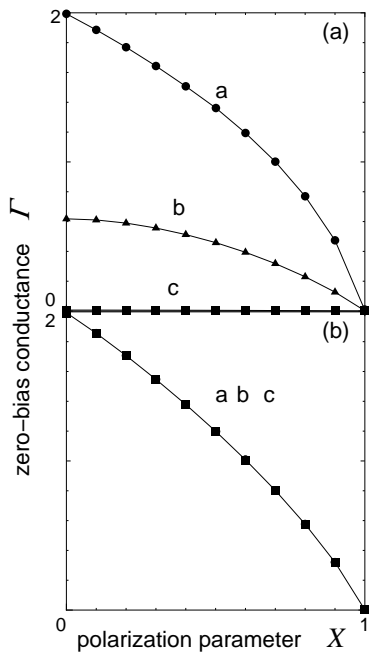}
}
\caption{
$X$ dependence of $\Gamma$ in SCF calculation for $d$-wave junctions. 
(a) $\alpha =0$ and (b) $\alpha =\pi /4$.
a: $Z=0$, b: $Z=1$ and c: $Z=5$. }
\end{center}
\end{figure}

\begin{figure}[htb]
\begin{center}
\scalebox{0.4}{
\includegraphics[width=15.0cm,clip]{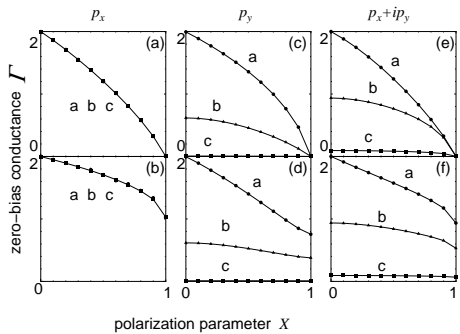}
}
\caption{
$X$ dependence of $\Gamma$ in SCF calculation. 
(a) $\theta_M =0$ and (b) $\theta_M =\pi /2$ 
with $p_x$-wave junctions.
(c) $\theta_M =0$ and (d) $\theta_M =\pi /2$ 
with $p_y$-wave junctions. 
(e) $\theta_M =0$ and (f) $\theta_M =\pi /2$ 
with $p_{x}+ip_y$-wave junctions. 
a: $Z=0$, b: $Z=1$ and c: $Z=5$. } 
\end{center}
\end{figure}

Finally, $X$ dependence  of $\Gamma$ is plotted for various temepratures. 
As seen from Fig. 2(b), using tunneling through ZES, we can determine the 
magnitude of $X$ thtough the value of $\Gamma$. 
This is a unique propetry for $d$-wave superconductor with $\alpha=\pi/4$, 
or $p$-wave junctions where all quasiparticles feel ABS independent of their 
directions of motions. 
Since $\Gamma$ is plotted at zero temperature in Fig. 2(b), it is 
actually important how this property holds even in finite temepratures. 
As shown in Fig. 6, when the magnitude of the 
temperature $T$ is sufficiently smaller than $T_{C}$, 
$\Gamma$ is almost insensitive to the magnitude of $Z$ 
and we can estimate the magnitude of $X$ through $\Gamma$. 
In the actual experiments, high $T_{C}$ 
cuprates, $e.g.$, YBaCuO, BiSrCaCuO, with (110) oriented 
interface is a promising candidate. 
In such a case, the actual value of $0.001T_{C}$ becomes 0.1K and 
this temperature is fully accesible in the experiments.

\begin{figure}[htb]
\begin{center}
\scalebox{0.4}{
\includegraphics[width=10.0cm,clip]{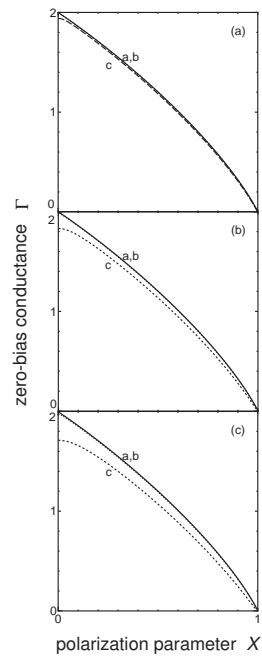}
}
\caption{
$X$ dependence of $\Gamma$ for various temperatures 
for $d$-wave junctions for $\alpha=\pi/4$ in non SCF calculation. 
(a) $T/T_{C}=0.001$, (b) $T/T_{C}=0.005$ 
and (c) $T/T_{C}=0.01$. a: $Z=0$, b: $Z=1$ and c: $Z=5$. } 
\end{center}
\end{figure}
\

\subsection{Temperature dependence of zero-bias conductance}

In this subsection, we discuss the temperature dependence of 
$\Gamma$.
In the first part, we show the 
conductance in the non-SCF calculation. Then the results are
compared with those in the SCF calculation in the second part.
At first let us focus on 
$d$-wave junctions 
for $\alpha =0$ as shown in 
Figs.~7 (a), (b), (c),
where the conductance is plotted as a function of temperatures for several
magnitudes of the exchange potential $X$. 
We note in these junctions that the ZES is not 
formed at the interface. 
For $Z=0$ (see Fig. 7(a)), the exchange potential in ferromagnets
significantly affects the temperature-dependence 
of $\Gamma$. 
For large $X$,
$\Gamma$ increases with the increase of $T$ as shown in the curve $d$ in 
Fig.~7(a). 
The Andreev reflection (two-electron process) is suppressed by the large
exchange potential and 
the current is mainly carried by single electron process.
While for small $X$,  
$\Gamma$ decreases with the increase of $T$ as shown in curve $a$ 
in Fig.~7(a). 
This is because the current at the zero-voltage 
is mainly carried by two-electron process through the Andreev reflection 
and the amplitude of the Andreev reflection is 
suppressed for $T \rightarrow T_{C}$.  
For $Z=5$ (Fig. 7(c)),  
$\Gamma$ becomes small around $T\sim 0$ 
because the insulating barrier suppresses the Andreev reflection. 
The results show that $\Gamma$ increases monotonically with increasing temperatures 
independent of $X$ 
since the current is mainly carried by single electron process.
For $Z=1$ (Fig. 7(b)), excepting for large $X$, 
the magnitude of $\Gamma$ has a non-monotonic temperature dependence, 
since the amplitude of single-electron process (Andreev reflection) is 
enhanced (suppressed) with the increase of $T$.  
\begin{figure}[htb]
\begin{center}
\scalebox{0.4}{
\includegraphics[width=15.0cm,clip]{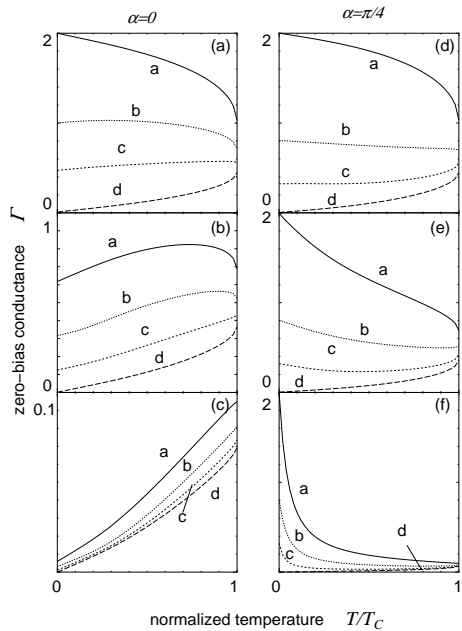}
}
\caption{
Temperature dependence of  $\Gamma$
in non-SCF calculation for $d$-wave junctions.
(a) $Z=0$, (b) $Z=1$ and (c) $Z=5$ with $\alpha =0$. 
(d) $Z=0$, (e) $Z=1$ and (f) $Z=5$ with  $\alpha =\pi /4$.  
 a: $X=0$, b: $X=0.7$, c: $X=0.9$ and d: $X=0.999$.} 
\end{center}
\end{figure}
\par

Secondly we show the temperature dependence of $\Gamma$
in $d$-wave junctions with $\alpha =\pi /4$ in 
Figs.~7(d), (e), and (f).
For $Z=0$, the line shape of the all curves in Fig.~7(d) are 
qualitatively similar to those with $\alpha =0$ shown in Fig.~7(a)
since ZES is not formed at the interface.
We note in Figs.~7 (d), (e) and (f) that $\Gamma$ at the zero temperature 
is independent of $Z$. 
The Andreev reflection is perfect in the limit of $Z=0$.
For finite $Z$, the ZES is formed at the interface which also leads to 
the perfect Andreev reflection at $T=0$.
The characteristic behavior of the resonant tunneling can be seen 
in the conductance for large $Z$. The amplitude of 
$\Gamma$ is proportional to the inverse of $T$ 
for intermediate temperatures (curve $a$ in Fig. 7 (f)). 
Since the retro-reflectivity of Andreev reflection is broken
by the exchange potentials, 
the degree of the resonance at the interface is weakened.
As a results, the temperature dependence deviates from the 
inverse of $T$ in curbs b, c and d in Fig.~7 (f). 
For sufficiently large magnitudes of $X$, $\Gamma$ becomes an 
increasing function of $T$ for $T>0.5T_C$ as shown in Fig. 7(f).  
In the case of the $d$-wave with $\alpha=\pi/4$, we can estimate the 
magnitude of $X$ from the temperature dependence of $\Gamma$. 

Thirdly we show temperature dependence of $\Gamma$ 
in $p_{x}$-wave junctions with  
$\theta_M =0$ in Figs. 8(a), (b), (c) 
and those with $\pi /2$ in Figs. 9(a), (b), (c). 
As shown in Figs.~8(a), (b) (c), the temperature dependences of 
$\Gamma$ are very similar to those of 
$d$-wave junctions with $\alpha =\pi /4$.
When $\theta_{M}=0$, 
$\Gamma$ for small $X$ is
a decreasing function of $T$, whereas $\Gamma$ for large $X$
is an increasing function of $T$. 
In the case of $\theta_{M}=\pi/2$, however,
$\Gamma$ becomes a monotonic decreasing function 
of $T$ independent of $Z$ as shown in  Figs. 9(a), (b), (c). 
The spin of a quasiparticle is always conserved in the Andreev reflection in this case. 
Therefore the suppression of the conductance 
due to the breakdown of the retro-reflectivity 
becomes weaker than that in the case of $\theta_{M}=0$.

Next we show the conductance in $p_{y}$-wave junctions with  
$\theta_M =0$ in Figs. 8(d), (e), (f) 
and those with $\pi /2$ in Figs. 9(d), (e), (f).  
In $\theta_{M}=0$, 
the temperature dependence of  
$\Gamma$ are similar to those of 
$d$-wave junctions with $\alpha =0$ shown in Figs.~7(a),(b),(c). 
In these junctions, no ZES  is expected at the interface.  
In $\theta_{M}=\pi/2$ as shown in 
Figs.~9(d), (e), (f), 
$\Gamma$ for large $X$ are larger 
than those with $\theta_{M}=0$ in low temperatures. 
In the case of $\theta_{M}=\pi/2$, the spin of a quasiparticle is conserved 
in the Andreev reflection, therefore, 
the suppression of conductance due to the breakdown of retro-reflectivity 
becomes weaker than that in the case of $\theta_{M}=0$. 

Finally we show the conductance 
in $p_{x}+ip_{y}$-wave junctions with  
$\theta_M =0$ in Figs. 8(g), (h), (i) 
and those with $\pi /2$ in Figs. 9(g), (h), (i).  
In $p_{x}+ip_{y}$-wave junctions with $\theta_{M}=0$,  
the temperature dependence of $\Gamma$ can be understood by the 
combination of the results in $p_{x}$-wave and those in $p_{y}$-wave junctions
because the ZES is only expected for a quasiparticle incident perpendicular 
 to the interface~\cite{Kusakabe}. 
As seen from Fig.~8 (g)
for $Z$=0, there is no clear difference between 
$\Gamma$ in $p_{x}+ip_{y}$-wave junctions with $\theta_{M}=0$
and corresponding results in $p_{x}$ or $p_{y}$-wave junctions shown in
 Figs.~8(a) and (d). 
For a finite barrier potential at $Z=1$, 
the line shape of the all curves in Fig.~8(h) 
is rather similar to corresponding results in the $p_{x}$-wave junctions 
than those in the $p_{y}$-wave junctions.
However, $\Gamma$ are smaller than those in the $p_{x}$-wave junctions.
For $Z=5$, $\Gamma$ for small $X$ are enhanced at low temperatures
because of the ZES [see curve $a$ in Fig. 8(i) ]. 
On the other hand, for large $X$, 
$\Gamma$ is an increasing function of $T$ 
[see curve $d$ in Fig.  8(i) ]. 
For $\theta_{M}=\pi/2$, $\Gamma$ becomes a
decreasing function of $T$ for all cases as shown in 
Figs. 9(g), (h), (i). These features are similar to those in the $p_{x}$-wave junctions. 

\begin{figure}[htb]
\begin{center}
\scalebox{0.4}{
\includegraphics[width=15.0cm,clip]{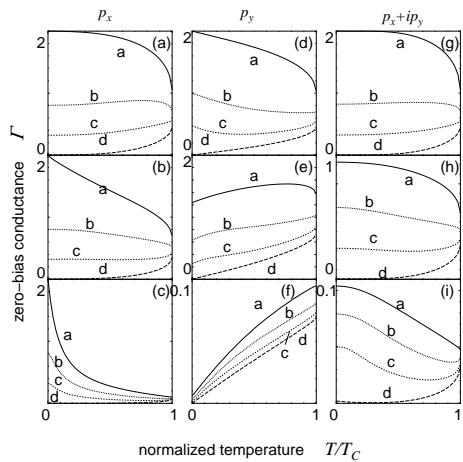}
}
\caption{
Temperature dependence of  $\Gamma$
in non-SCF calculation
for $\theta_{M} =0$.
(a) $Z=0$, (b) $Z=1$ and (c) $Z=5$ in $p_{x}$-wave junctions.
(d) $Z=0$, (e) $Z=1$ and (f) $Z=5$ in $p_{y}$-wave junctions.
(g) $Z=0$, (h) $Z=1$ and (i) $Z=5$ in $p_{x}+ip_{y}$-wave junctions.
 a: $X=0$, b: $X=0.7$, c: $X=0.9$ and d: $X=0.999$. 
}
\end{center}
\end{figure}

\begin{figure}[htb]
\begin{center}
\scalebox{0.4}{
\includegraphics[width=15.0cm,clip]{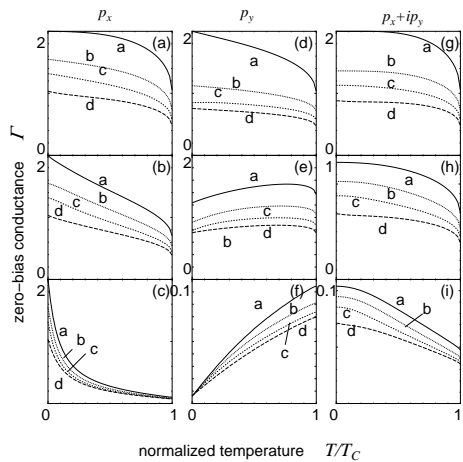}
}
\caption{
Temperature dependence of  $\Gamma$
in non-SCF calculation
for  $\theta_{M} =\pi/2$.
(a) $Z=0$, (b) $Z=1$ and (c) $Z=5$ in $p_{x}$-wave junctions.
(d) $Z=0$, (e) $Z=1$ and (f) $Z=5$ in $p_{y}$-wave junctions.
(g) $Z=0$, (h) $Z=1$ and (i) $Z=5$ in $p_{x}+ip_{y}$-wave junctions.
 a: $X=0$, b: $X=0.7$, c: $X=0.9$ and d: $X=0.999$. } 
\end{center}
\end{figure}

It is important to check how the above results 
are modified in the presence of the spatial dependence in the pair potential 
near the interface. We show the conductance in SCF calculation in 
the second part 
of this subsection.
In what follows, we consider two cases of $X$, 
$X=0$ (curve $a$) 
and $X=0.9$ (curve $b$). 
The corresponding results in non-SCF calculation 
are curves $a$ (X=0) and $c$ (X=0.9) from 
Figs.~7 to 9. 

In Fig.~10, we show the conductance in SCF calculation 
in the $d$-wave junctions with $\alpha=0$ and $\pi/4$.
It is found that the temperature dependences of $\Gamma$ with $\alpha=0$ in the SCF 
calculation are very similar to those in the non-SCF calculation 
when we compare the results in Figs.~7(a), (b), (c) with those in 
Figs.~10(a), (b), (c). 
The same tendency can be seen between the results 
the $d$-wave junctions with $\alpha=\pi/4$ in 
Figs.~7 (d), (e), (f) and those in Figs.~10(d), (e), (f). 
When the ZES are formed at the interface, the profile of the pair
potential significantly deviates from the step-function.
The characteristic behavior of the conductance in the SCF, however, is qualitatively
the same with those in the no-SCF. From the calculated 
results, we conclude that the conductance 
is insensitive to the profile of the pair potential.
This is because the resonant tunneling through the ZES dominates the conductance. 
We note that the ZES is a consequence of the sign-change of the pair potential.

\begin{figure}[htb]
\begin{center}
\scalebox{0.4}{
\includegraphics[width=15.0cm,clip]{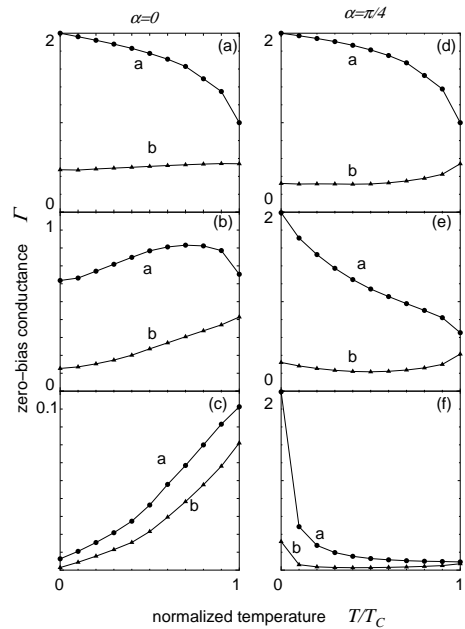}
}
\caption{
Temperature dependence of  $\Gamma$
in SCF calculation  for  $d$-wave junctions.
(a) $Z=0$, (b) $Z=1$ and (c) $Z=5$ with $\alpha =0$. 
(d) $Z=0$, (e) $Z=1$ and (f) $Z=5$ for $\alpha =\pi /4$.  
 a: $X=0$ and b: $X=0.9$. } 
\end{center}
\end{figure}
In the $p_{x}$-wave junctions, line shapes of 
$\Gamma$ for $\theta_M =0$ shown in Figs.~11(a), (b), (c) 
are very similar to those in the $d$-wave junctions 
with $\alpha=\pi/4$. 
For $\theta_M =\pi /2$,
as shown in Figs.~12(a), (b), (d), the magnitudes of 
$\Gamma$ are slightly larger than those in Figs.~11(a), (b), (c).
These features are almost similar to those 
found in non-SCF calculation in Figs.~8(a), (b), (c) and Figs.~9(a), (b), (c). 
\begin{figure}
\begin{center}
\scalebox{0.4}{
\includegraphics[width=15.0cm,clip]{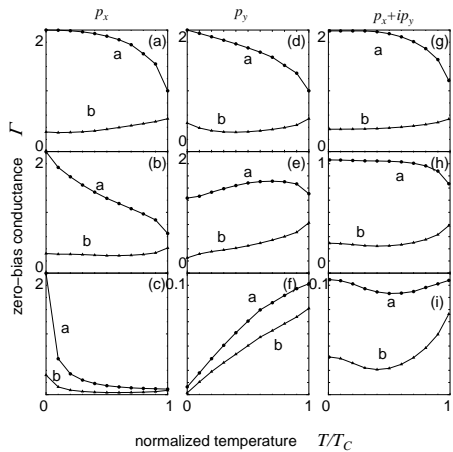}
}
\caption{
Temperature dependence of  $\Gamma$
in SCF calculation for  $\theta_{M} =0$.
(a) $Z=0$, (b) $Z=1$ and (c) $Z=5$ in $p_{x}$-wave junctions.
(d) $Z=0$, (e) $Z=1$ and (f) $Z=5$ in $p_{y}$-wave junctions.
(g) $Z=0$, (h) $Z=1$ and (i) $Z=5$ in $p_{x}+ip_{y}$-wave junctions.
 a: $X=0$ and b: $X=0.9$. } 
\end{center}
\end{figure}
\begin{figure}[htb]
\begin{center}
\scalebox{0.4}{
\includegraphics[width=15.0cm,clip]{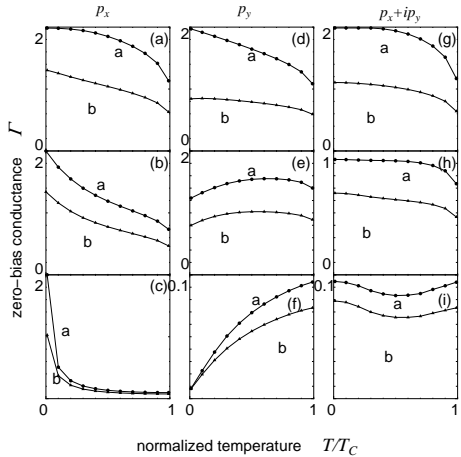}
}
\caption{
Temperature dependence of  $\Gamma$
in non-SCF calculation for  $\theta_{M} =\pi/2$.
(a) $Z=0$, (b) $Z=1$ and (c) $Z=5$ in $p_{x}$-wave junctions.
(d) $Z=0$, (e) $Z=1$ and (f) $Z=5$ in $p_{y}$-wave junctions.
(g) $Z=0$, (h) $Z=1$ and (i) $Z=5$ in $p_{x}+ip_{y}$-wave junctions.
 a: $X=0$ and b: $X=0.9$. } 
\end{center}
\end{figure}
As well as in the $p_x$-wave junctions, the characteristic behavior
of $\Gamma$ in SCF results in $p_{y}$-wave junctions shown in 
Figs.~11(d), (e), (f) and Figs.~12(d), (e), (f) 
are almost the same with those obtained in 
non-SCF calculation shown in 
Figs.~8(d), (e), (f) and Figs.~9(d), (e), (f).

In the case of $p_{x}+ip_{y}$-wave symmetry, 
the line-shapes of $\Gamma$ for (g), (h) and (i) in 
Figs.~11 and 12 are similar to those in 
non-SCF results shown in (g) and (h) in Figs.~8 and 9. 
However, the temperature dependencies of 
$\Gamma$ based on the SCF calculation deviate from those 
in the non-SCF one for large $Z$ [Fig. 11(i) and Fig. 12(i)]. 
In the SCF calculation, $\Gamma$ first decreases with the 
increase of $T$ then increases.  
The decreasing part is similar to that in the $p_{x}$-wave 
case and is explained by the ZES. The increasing part is similar to 
that in the $p_{y}$-wave junctions. 
Since only a quasiparticle injected perpendicular 
to the interface contributes to the ZES, the effects of the 
spatial dependence of the pair potentials on the
conductance are not be negligible~\cite{Yamashiro3}.  

\section{Summary}
In this paper, we have calculated the 
polarization and temperature dependence of the 
zero-bias conductance (ZBC) in F/I/S junctions, 
where we have chosen the symmetry of the pair potential
as $d$-wave for high $T_{C}$ cuprates 
and $p_{x}+ip_{y}$-wave  for Sr$_2$RuO$_4$.
As a reference, we have also studied the conductance in
$p_{x}$ and $p_{y}$-wave junctions.  
We have established a formalism of the ZBC which 
is available for the arbitrary $\theta_M$ which is  
the angle between the magnetization axis in ferromagnet 
and $c$-axis of superconductors. 
The $\theta_M$ dependence
of the tunneling conductance only appears in the spin-triplet superconductor junctions.
On the basis of the numerical results, we reach the following conclusions. \par
\noindent
(1) When injected quasiparticles from ferromagnets 
always feel the zero-energy resonance state, 
$e.g.$, $d$-wave junction with (110) orientation and
$p_{x}$-wave junction,  
the zero-bias tunneling conductance (ZBC) at the zero temperature 
is insensitive to the barrier potential at the interface. 
This property is useful for the determination of the 
degree of the spin-polarization in ferromagnets at sufficiently 
low temperatures.  
One of the promising candidate is 
LaSrMnO/YBaCuO(BiSrCaCuO)  with well oriented (110) oriented 
interface. 
Within our theory, below 0.1K 
we can estimate the magnitude of polarization of 
ferromagnets
through the value of the conductance of the junctions. \par
\noindent
(2) For $d$-wave junctions with (110) orientation and
$p_{x}$-wave junctions, the ZBC decreases with increasing temperatures
 when the degree of the polarization, $X$, is small.
For large $X$, the ZBC is an increasing function of $T$. 
The presence of the ZES explains these behavior.
In $d$-wave junctions with (100) orientation and
$p_{y}$-wave junctions, the ZBC is an increasing function of $T$ independent of $X$. 
This is because the ZES does not appear at these junction interface.
For $p_{x}+ip_{y}$-wave junctions, the 
ZBC first decreases with increasing $T$ then increases. \par
\noindent
(3) In $p$-wave junctions, the temperature dependence of the ZBC
depends on the direction of the magnetization axis of 
ferromagnets because of the spin degree of freedom of Cooper pairs
in spin-triplet superconductors.  \par
\noindent
(4)
Throughout this paper, we have calculated the ZBC in two ways; 
i) the spatial dependence of the pair potential is 
assumed to be the step function (non-SCF calculation), 
ii) the spatial depletion of the pair potentials are determined 
self-consistently (SCF-calculation). 
We have confirmed that there are no remarkable differences between the conductance 
in the non-SCF calculation and those in the SCF calculation.
\par 
In this paper, effects of random potentials in 
ferromagnets are not taken into account. 
Recently, there are several works on 
random scattering effects in unconventional 
superconductor junctions~\cite{Itoh2002,Asano2000,Nazarov,YY,future}. 
It is actually interesting to 
study transport properties of junctions where diffusive ferromagnets are attached to unconventional superconductors. 
In the present paper, the splitting of the ZBCP by magnetic fields 
through the Zeeman effect or magnetic impurities in an insulator are not taken into account~\cite{Zhu,Kashiwaya4,Zutic}. 
These are interesting and important future issues.

\acknowledgements 

This work was partially supported by the Core Research for
Evolutional Science and Technology (CREST) of the Japan Science and
Technology Corporation (JST).
J.I. acknowledges support by the NEDO 
international Joint Research project 
"Nano-scale Magnetoelectronics". 


%

\begin{thebibliography}{}
\bibitem{future}
The present model deals with the ideal situation where the height of 
the ZBCP is only the function of the polarization of the ferromagnet. 
However, in real junctions, the interface quality and the impurity 
distribution near the interface may have influences on the ZBCP height. 
Detailed analysis of such factors will be discussed in coming papers.
\end{thebibliography}
\end{document}